\documentclass[journal,10pt]{IEEEtran}
\usepackage{amsmath,amssymb,amsfonts,mathrsfs,bm}
\usepackage{amstext}
\usepackage{upgreek}
\usepackage{multicol}
\usepackage{graphicx}
\usepackage{paralist}
\usepackage{cite}
\usepackage{citesort}
\usepackage{booktabs}
\usepackage{multirow}
\usepackage{subfigure}
\newtheorem{proposition}{Proposition}

\IEEEoverridecommandlockouts

\begin{document}

\title{Optimal Caching and Scheduling for Cache-enabled D2D Communications}
\author{
\IEEEauthorblockN{{Binqiang Chen, Chenyang Yang and Zixiang Xiong}}
\vspace{-3mm}
\thanks{
	B. Chen and C. Yang are with Beihang University, Beijing, China, Emails: \{chenbq, cyyang\}@buaa.edu.cn. Z. Xiong is with Texas A\&M University, College Station, TX 77843, USA, E-mail: zx@ece.tamu.edu. This work is supported by NSF Grants with No. 61671036, 61301085 and 61429101.
	}\vspace{-3mm}
%
}
\maketitle

\begin{abstract}
To maximize offloading gain of cache-enabled device-to-device (D2D) communications, content placement and delivery should be jointly designed. In this letter, we jointly optimize caching and scheduling policies to maximize successful offloading probability, defined as the probability that a user can obtain desired file in local cache or via D2D link with data rate larger than a given threshold. We obtain the optimal scheduling factor for a random scheduling policy that can control interference in a distributed manner, and a low complexity solution to compute  caching distribution. We show that the offloading gain can be remarkably improved by the joint optimization.
\end{abstract}

\vspace{-1mm}
\vspace{-5mm}
\section{Introduction}
\vspace{-1mm}
\label{s:1}
Cache-enabled device-to-device (D2D) is a promising way to offload data traffic, especially video, of cellular networks \cite{Golrezaei.TWC,JMY.JSAC}, which attracts considerable attention recently  \cite{afshang2015fundamentals,shankar2016effect,Chedia2016effect,Plac.D2D,mehrnaz2016optimal,zhang2016efficient}.


In \cite{afshang2015fundamentals}, the impact of clustered user location on coverage probability of cache-enabled D2D network, defined as the probability that a randomly located user has signal to interference and noise ratio greater than a threshold, was analyzed. In \cite{shankar2016effect,Chedia2016effect},  the impact of user mobility on cache-enabled D2D were investigated, where the performance metrics are respectively coverage probability and service
success probability, defined as the probability that the desired file of a user can be transmitted completely within the exponentially distributed lifespan of a D2D link. In \cite{JMY.JSAC,Plac.D2D,mehrnaz2016optimal}, probabilistic caching policies were optimized for cache-enabled D2D, where cache hit probability was maximized in \cite{JMY.JSAC}, which reflects the probability that a user can find the requested file from other users in proximity, and coverage probability was maximized in  \cite{Plac.D2D,mehrnaz2016optimal}.
However, the resulting optimal caching policies cannot satisfy the data rate requirement of each user \cite{singh2013offloading},
which is critical to support the traffic with quality of service such as video streaming. In  \cite{zhang2016efficient}, scheduling and power allocation was optimized for cache-enabled D2D considering a given probabilistic caching policy.
In cache-enabled wireless networks, interference has a large impact on the offloading gain and
the policy optimization. This implies that the joint design of content delivery and content placement is
important, which however has not been addressed in existing works for cache-enabled D2D.

Stochastic geometry \cite{singh2013offloading} is a popular tool to analyze wireless networks with local caching \cite{afshang2015fundamentals,Chedia2016effect,shankar2016effect,Plac.D2D,mehrnaz2016optimal,zhang2016efficient,B2015optimal}. Different from caching at the BS \cite{B2015optimal}, where all BSs generate interference,
in real world D2D networks, the users without establishing D2D links will not transmit and hence do
not generate interference. However, priori works using stochastic geometry theory for cache-enabled
D2D either assume that all users in the network generate interference  \cite{Plac.D2D},
or assumes that a given fraction of the users in the network generate interference \cite{mehrnaz2016optimal}. Such
assumptions give rise to convex optimization for caching policy, but lead to pessimistic or inaccurate characterization of
the interference level. It is worth to note that the framework in  \cite{mehrnaz2016optimal} is general and is not restricted to D2D.

In this letter, we strive to jointly optimize caching and scheduling policies for cache-enabled D2D communications to maximize successful offloading probability, defined as the probability that a user can obtain desired file in its local cache or via a D2D link with data rate higher than a predetermined threshold.
We consider probabilistic caching policy, which has been widely considered for caching in wireless edge \cite{B2015optimal,JMY.JSAC,Plac.D2D,mehrnaz2016optimal}, and introduce a random scheduling policy to control interference by a scheduling factor, both can be implemented in a distributed manner.
We first derive the closed-form expression of the successful offloading probability. Since we do not assume that all links generate interference as in \cite{B2015optimal} or only a fixed fraction of users generate interference as in \cite{mehrnaz2016optimal}, the resulting caching problem is non-convex. After obtain the closed-form expression of the optimal scheduling factor, we find the local optimal caching distribution via interior point method. To obtain a low complexity caching policy, we maximize a lower bound of the successful offloading probability for large collaboration distance. Simulations show that joint optimization is vital to improve the offloading gain.
\vspace{-8mm}
\section{System Model}
\vspace{-1mm}
\label{s:2}
Consider a cell where users' locations follow a Poisson point process (PPP) with density $\lambda$.
Each single-antenna user has a local cache to store files, and can act as a helper  to share files.
For simplicity, we assume that each user can only store one file in its local cache as in \cite{Golrezaei.TWC,shankar2016effect,Plac.D2D}.
\vspace{-5mm}
\subsection{Content Popularity and Caching Placement}
\vspace{-1mm}
Consider a static content catalog consisting of $N_f$ files that all users in the cell may request, which are
indexed in descending order of popularity. The probability that the $i$th file is requested follows a Zipf distribution
$q_i={i^{-\beta}}/{\sum_{k=1}^{N_f}k^{-\beta}}$,
%
where $\sum_{i=1}^{N_f} q_i=1$, and the parameter $\beta$ reflects how skewed the popularity distribution is \cite{Zipf99}.

Consider a probabilistic caching policy where each user independently selects a file to cache according to a specific probability distribution $\textbf{c} = \{c_1, c_2, ... , c_{N_f}\}$, where $c_i$ is the probability that a user caches the $i$th file. According to the thinning property \cite{PPP}, the locations of users cached the $i$th file follow a PPP with density $\lambda_i = \lambda c_i$.

%
\vspace{-5mm}
\subsection{Content Delivery and Random Scheduling}
\vspace{-1mm}
If a user can find the desired file in its own cache, the user directly retrieves the file. If a user can find its requested file in the local caches of helpers in proximity, it fetches the file via D2D link. Otherwise, the user accesses to the BS.

Before the file is delivered by a helper, a D2D link needs to be established with the user who requests a file, called a D2D receiver (DR).
After a helper establishes a D2D link, the helper acts as a D2D transmitter (DT). For any DR, only the nearest helper from those with distances smaller than a given \emph{collaboration distance} $r_{\rm c}$ serves as a DT. The D2D link between a DT and a DR does not change during the file delivery. Assume that the user can simultaneously transmit and receive as in  \cite{Plac.D2D,mehrnaz2016optimal}. Assume that a fixed bandwidth of $W$ is assigned to the D2D links to avoid mutual interference with cellular links, and all DTs transmit with the same transmit power $P_t$. The BS is aware of the files cached at the users and helps establish and synchronize the D2D links.
\vspace{-4mm}
\begin{figure}[!htb]
	\centering
	\includegraphics[width=0.45\textwidth]{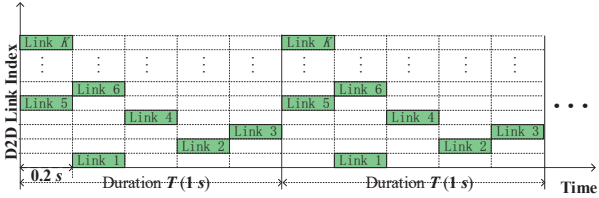}\vspace{-3mm}
	\caption{Illustration of the random scheduling policy for established D2D links, where $\tau=0.2$, the duration of each period $T=1$ second. Then, each period can be divided into five time slots.}\label{fig.0}
\vspace{-3mm}
\end{figure}

To control interference among the established D2D links in a distributed manner, we consider a random scheduling policy, which operates periodically with a given duration $T$, as illustrated in Fig. \ref{fig.0}. The value of $T$ can be selected to control the quality of service of multiple DRs, e.g., to trade off the initial delay and stalling for video streaming. We divide $T$ into multiple synchronized time slots each with duration $\tau T$. In each period, each DT independently chooses a time slot to transmit and stays muting in the remaining time of the period. We call $\tau \in \left(0,1\right]$ as \emph{scheduling factor}. When  there are $K$ DTs, if  $\tau=1/K$ and only one DT is allowed to transmit in each time slot, then the scheduling degenerates to time division multiple access. When $\tau=1$, all established D2D links transmit simultaneously without any interference control, which is considered in  \cite{afshang2015fundamentals,Plac.D2D,mehrnaz2016optimal}.

Consider the interference-limited regime and hence neglect the noise. The signal to interference ratio (SIR) at the DR requesting the $i$th file is
$\gamma_{i,r}  = \frac{P_thr^{-\alpha}}{\sum_{k \in \mathcal{U}_\tau}  P_t h_k r_k^{-\alpha}} = \frac{hr^{-\alpha}}{ I_{i,r}}$,
where $h$ is the channel power gain that follows an exponential distribution with unit mean for Rayleigh fading, $r$ is the D2D link distance, $\alpha$ is the path loss exponent, $I_{i,r} =  \sum_{k \in \mathcal{U}_\tau} h_k r_k^{-\alpha} $ is the total interference from all the other DTs transmitting at the same time slot (constituting the set $\mathcal{U}_\tau$) normalized by $P_t$. Then, the data rate of the DR requesting the $i$th file with D2D link distance $r$ is
$R_{i,r}  = \tau W\log_2\left(1+\frac{hr^{-\alpha}}{I_{i,r}}\right)$.

With the growth of $\tau$, the number of D2D links simultaneous transmitting in each time slot increases, leading to more interference, meanwhile the resource allocated to each link increases. Hence, the rate can be controlled  by adjusting $\tau$. \vspace{-3mm}

\section{Optimal Caching and Scheduling Policies}
\label{s:4}
In this section, we optimize the caching policy and scheduling policy. To this end, we first derive the successful offloading probability to reflect the offloading gain. Then, we jointly optimize the caching distribution and scheduling factor to maximize the successful offloading probability.
\vspace{-4mm}

\subsection{Successful Offloading Probability}
\label{s:4.1}
\vspace{-0mm}
Define the successful offloading probability as the probability that a user can find its requested file in own cache, or in caches of helpers with distance $r \le r_{\rm c}$ and with D2D link data rate higher than a required threshold $R_0$.

The probability density function (pdf) of the distance between a user requesting the $i$th file and its nearest helper cached the $i$th file is $f_i(r)=2\pi r \lambda_i e^{-\lambda_i \pi r^2}$ \cite{PPP}. Then, the successful offloading probability is
\vspace{-1mm}
\begin{equation}\label{equ_p_s_define}\textstyle
\mathbb{P}_{\text{o}}(\textbf{c},\tau) =\sum_{i=1}^{N_f} q_i \left( c_i+ \left(1-c_i\right) \int_{0}^{r_c} f_i(r)  \mathbb{P}\left[R_{i,r}>R_0\right] dr\right).
\end{equation}

\begin{proposition}\label{t:1}
The successful offloading probability can be approximated as
\vspace{-3mm}
\begin{equation}
\label{equ.t:1.1}\textstyle
\mathbb{P}_{\text{o}}(\textbf{c},\tau) \approx \sum_{i=1}^{N_f}q_i \left(c_i +  \frac{(1-c_i)\lambda_i}{A_i} ( 1-e^{-\pi A_ir_{\rm c}^2 })  \right) ,\vspace{-2mm}
\end{equation}
where $A_i \triangleq \lambda_i + \tau \lambda_0 \xi \gamma_0^{2/\alpha}$, $\gamma_0 \triangleq e^{\frac{R_0\ln2}{\tau W}} - 1$, $\xi \triangleq \int_{0}^{+\infty} \frac{1}{1+t^{\alpha/2}}dt$, $\lambda_0$ is the density of all DTs with expression \vspace{-2mm}
\begin{equation}\textstyle\label{equ.lambda_I}
	 \lambda_0=    \sum_{i=1}^{N_f} \lambda_i  \left(1 - \left(1+\frac{ q_i}{3.5 c_i} \right)^{-3.5} \theta_i \right),\vspace{-2mm}
\end{equation}
$\theta_i \triangleq \frac{\Gamma\left(3.5,\left(3.5\lambda_i+\lambda q_i\right) \pi r_{\rm c}^2\right)}{\Gamma(3.5,3.5\lambda_i \pi r_{\rm c}^2)}$, and $\Gamma(s,x) = \int_{0}^{x}t^{s-1}e^{-t}dt$ is the lower incomplete gamma function.
\end{proposition}

\begin{IEEEproof}
The probability that the data rate of a D2D link  with distance $r$ to transmit the $i$th file is larger than $R_0$ is
\vspace{-1mm}\begin{equation}
\begin{aligned} \textstyle
\mathbb{P}\left[R_{i,r}>R_0\right]
& \textstyle \stackrel{(a)}{=} \mathbb{P} \left[\frac{hr^{-\alpha}}{I_{i,r}} \geq \gamma_0 \right]  =\mathbb{P} \left[ h \geq \gamma_0 r^{\alpha}I_{i,r}\right] \\
& \textstyle\stackrel{(b)}{=} \mathbb{E}_{I_{i,r}} \left[ \exp\left( -r^{\alpha}\gamma_0 I_{i,r}  \right)  \right] =   \mathcal{L}_{I_{i,r}}\left(r^{\alpha}\gamma_0 \right),\nonumber
\end{aligned}
\end{equation}
where (a) is obtained from expressions of $R_{i,r}$ and $\gamma_0$, (b) is by the fact that $h$ follows an exponential distribution, and $\mathcal{L}_{I_{i,r}}(s)$ is the Laplace transform of the random variable $I_{i,r}$.

To derive $\mathcal{L}_{I_{i,r}}(s)$, we need to obtain the density of DTs cached the $i$th file since not all helpers act as DTs. Since it is hard to directly derive the probability that a helper cached the $i$th file acts as a DT, denoted as $p_{{\rm a},i}$, we first derive its complementary probability $p_{{\rm o},i}$, which is the probability that no DR requesting the $i$th file is accessed to the helper. Considering that a DR can access to a helper only if their distance is less than $r_c$, then $p_{{\rm o},i}$ is obtained as
\vspace{-2mm}\begin{equation}
\label{equ.pa}
\begin{aligned}
\textstyle
p_{{\rm o},i} &\textstyle= \frac{\int_{0}^{\pi r_c^2} e^{-\lambda q_i x}g_x(x,\lambda_i) dx}{\int_{0}^{\pi r_c^2} g_x(x,\lambda_i) dx}  =  \left(1+\frac{ q_i}{3.5 c_i}\right)^{-3.5}\theta_i,
\end{aligned}\vspace{-2mm}
\end{equation}
where $g_x(x,\lambda_i) =  \frac{3.5^{3.5}}{\Gamma(3.5,\infty)}\lambda_i^{3.5}x^{2.5}e^{-3.5\lambda_ix}$ is the pdf of the coverage area $x$ for a typical Voronoi cell \cite{PPPactive}. Then, $p_{{\rm a},i} = 1- p_{{\rm o},i}$. According to the thinning property, the DTs cached the $i$th file follow a PPP with density $\lambda^d_i=\lambda_i p_{{\rm a},i}$, and the density of all DTs is $\lambda_0 = \sum_{i=1}^{N_f}\lambda_i^d$.
Considering the random scheduling policy,
the DTs cached with the $i$th file transmit in a time slot are with density $\tau \lambda^d_i$, whereas the DTs cached with other files who transmitting in the time slot are with density $\tau(\lambda_0-\lambda^d_i)$ and could be closer than the desired DT. Using Theorem 2 in \cite{PPP}, when $\alpha>2$ we can derive that
\vspace{-2mm}\begin{equation}
\label{pr:1.1}
\begin{aligned}
& \textstyle\mathcal{L}_{I_{i,r}}(r^{\alpha}\gamma_0)  = \exp\left(-2\pi\tau \left(\lambda_0-\lambda_i^d \right) \int_{0}^{ \infty} \frac{\gamma_0}{\gamma_0+(v/r)^{\alpha}} v dv \right. \\
&\textstyle \left. -2\pi\tau \lambda_i^d \int_{r}^{ \infty} \frac{\gamma_0}{\gamma_0+(v/r)^{\alpha}} v dv   \right)\\
& \textstyle  \stackrel{(a)}{=} \exp(-\pi r^2 \tau \gamma_0^{2/\alpha}(\lambda_0 \int_{0}^{ \infty} \frac{1}{1+u^{\alpha/2}} du
+ \lambda_i^d \int_{0}^{\gamma_0^{-2/\alpha}} \frac{1}{1+u^{\alpha/2}} du)   ) \\
&\textstyle \approx \exp(-\pi r^2  \tau \lambda_0 \xi \gamma_0^{2/\alpha} ),
\end{aligned}
\end{equation}
where (a) is obtained by employing a change of variable $u=({v}/({r\gamma_0^{1/\alpha}}))^2$. The approximation comes from the fact that $\xi = \int_{0}^{ \infty} \frac{1}{1+u^{\alpha/2}} du >\int_{0}^{\gamma_0^{-2/\alpha}} \frac{1}{1+u^{\alpha/2}} du$ and $\lambda_0 = \sum_{k=1}^{N_f}\lambda_k^d  \gg \lambda_i^d$, which is accurate when $N_f$ is large and caching distribution \textbf{c} is not too skewed. By substituting \eqref{pr:1.1} into \eqref{equ_p_s_define}, we obtain \eqref{equ.t:1.1}.
\end{IEEEproof}
In Proposition \ref{t:1}, $\gamma_0$ is the SIR threshold corresponding to the data rate threshold $R_0$, while $\tau \lambda_0$ is the density of DTs simultaneously transmit in a time slot and reflects the interference level. By considering that some users do not generate interference (i.e., $\lambda_0 < \lambda$), the offloading gain can be computed more accurate than \cite{afshang2015fundamentals,Plac.D2D,mehrnaz2016optimal}.
%

\vspace{-3mm}
\subsection{Optimal Caching Policy and Random Scheduling Policy}
\vspace{-1mm}
To maximize the offloading gain introduced by cache-
enabled D2D communications, we jointly optimize the caching and scheduling policies by solving the following problem,
\vspace{-2mm}\begin{subequations}
\label{equ.opt}
\begin{align}
\textbf{P1}: \quad& \max_{\tau,\textbf{c}}\,\, && \textstyle \mathbb{P}_{\text{o}}(\textbf{c},\tau) \nonumber\\
&  \textstyle s.t.  && \textstyle \sum_{i=1}^{N_f} c_i=1,  \textstyle 0\leq c_i \leq 1, i=1,...,N_f, \label{equ.opt.a}\\
& && 0< \tau \leq 1 \label{equ.opt.b}.
\end{align}
\end{subequations}\vspace{-2mm}

\vspace{-2mm}
\begin{proposition}\label{t:2} For any given caching distribution $\textbf{c}$, The optimal scheduling factor is
	\vspace{-1mm}\begin{equation}
	\label{equ.sp:1.1}\textstyle
	\tau^* = 1/\lceil \frac{W \kappa}{R_0 \ln 2} \rceil,\vspace{-1mm}
	\end{equation}
where $\lceil x \rceil$ is the smallest integer greater than or equal to $x$, $\kappa = \mathcal{W}(-\frac{\alpha}{2}e^{-\frac{\alpha}{2}}) + \frac{\alpha}{2}$, and $\mathcal{W}(\cdot)$ is the Lambert-W function. 
\end{proposition}
\begin{IEEEproof}
The derivative of $\mathbb{P}_{\text{o}}(\textbf{c},\tau)$ with respect to $\tau$ is
\vspace{-2mm}\begin{equation}
	\begin{aligned}\textstyle
	\frac{\partial \mathbb{P}_{\text{o}}(\textbf{c},\tau) }{\partial \tau}
	&\textstyle= \sum_{i=1}^{N_f} q_i (1-c_i)\lambda_i \frac{\partial A_i}{\partial \tau} \frac{e^{-\pi A_i r_c^2}(1+\pi A_i r_c^2)-1}{A_i^2}\\
	&\textstyle \stackrel{(a)}{=} -g(\tau) \sum_{i=1}^{N_f} q_i (1-c_i)\lambda_i \frac{e^{-\pi A_i r_c^2}(1+\pi A_i r_c^2)-1}{A_i^2},
	\end{aligned}\vspace{-2mm}
\end{equation}
where (a) is because $g(\tau) \triangleq \frac{\partial A_i}{\partial \tau} = \frac{\lambda_0\xi}{\tau}(e^{\frac{R_0\ln2}{W\tau}}-1)^{\frac{2}{\alpha}-1}((\tau-\frac{2R_0\ln2}{W\alpha})e^{\frac{R_0\ln2}{\tau W}}-\tau)$. It is not hard to show that $g(\tau)$ is an increasing function of $\tau$. Since $e^{x}>(1+x)$ when $x>0$, $\frac{e^{-\pi A_i r_c^2}(1+\pi A_i r_c^2)-1}{A_i^2}<0$ always holds. Therefore, the optimal scheduling factor can be obtained from $g(\tau)=0$. Since the equation $e^{ax+b}=cx+d$ can be expressed as $x= -\frac{1}{a}\mathcal{W}(-\frac{a}{c}e^{b-\frac{ad}{c}})-\frac{d}{c}$ and $1/\tau^*$ should be an integer, we can obtain \eqref{equ.sp:1.1}.
\end{IEEEproof}

We can observe that $\tau^*$ depends on $R_0$, $W$ and $\alpha$ and is independent from caching distribution, but the optimal caching policy depends on $\tau^*$. Therefore, the optimal $\textbf{c}$ can be found by maximizing $\mathbb{P}_{\text{o}}(\textbf{c},\tau^*)$  under constraint \eqref{equ.opt.a} (called problem {\bf P2}). However, the objective function $\mathbb{P}_{\text{o}}(\textbf{c},\tau^*)$ is not concave in $\textbf{c}$ due to the coupling terms $c_i$ in the complicated expression $\lambda_0$ shown in \eqref{equ.lambda_I}. We can use the interior point method to obtain local optimal solution of $\textbf{c}$, which depends on the initial value. Nonetheless, we can increase the probability to find the global optimal solution of problem {\bf P2} by using the interior point method with multiple random initial values and then picking the solution with highest successful offloading probability.

\vspace{-4mm}
\subsection{Low Complexity Caching Policy}
\vspace{-1mm}
To obtain a low complexity solution, we consider $r_c \rightarrow \infty$, then from \eqref{equ.t:1.1} the successful offloading probability becomes
\vspace{-1mm}
\begin{equation}
	\label{equ.co.1.1}\textstyle
		\mathbb{P}^{\infty}_{\text{o}}(\textbf{c},\tau) = \sum_{i=1}^{N_f} q_i \left(c_i + \frac{(1-c_i)c_i}{c_i + \tau p_a \xi \gamma_0^{2/\alpha}} \right),
\end{equation}
where $p_a =  \sum_{i=1}^{N_f} c_i (1 - (1+{ q_i}/{3.5c_i} )^{-3.5} )$ is the probability that a helper acts as a DT. In what follows we find the caching distribution to maximize a lower bound of ${\mathbb{P}}^{\infty}_{\text{o}}(\textbf{c},\tau^*) $ (denoted as $\underline{\mathbb{P}}^{\infty}_{\text{o}}(\textbf{c},\tau^*) $). In Section \ref{s:5}, we will show the impact of $r_c$.

Considering that $p_a$ makes $\mathbb{P}^{\infty}_{\text{o}}(\textbf{c},\tau^*)$ in \eqref{equ.co.1.1} non-concave in $c_i$, we introduce a  $c_i-$independent upper bound of $p_a$  (denoted as $\bar{p}_a$ ). It is not hard to show that $p_a$ is concave in $\textbf{c}$, thus $\bar{p}_a$ can be obtained as the maximal ${p}_a$ by optimizing $c_i$. Because $\mathbb{P}^{\infty}_{\text{o}}(\textbf{c},\tau^*)$ decreases as  $p_a$ increases as shown in \eqref{equ.co.1.1}, $\bar{p}_a$ yields a lower bound of $\mathbb{P}^{\infty}_{\text{o}}(\textbf{c},\tau^*)$ as
\vspace{-2mm}\begin{equation}
\label{equ_p_s_l}
\begin{aligned}\textstyle
\underline{\mathbb{P}}^{\infty}_{\text{o}}(\textbf{c},\tau^*) = \sum_{i=1}^{N_f} q_i \left(c_i + \frac{(1-c_i)c_i}{c_i + \tau^* \bar{p}_a \xi \gamma_0^{2/\alpha}} \right).
\end{aligned}\vspace{-2mm}
\end{equation}

\begin{proposition}\label{t:3} The optimal caching probability to maximize $\underline{\mathbb{P}}^{\infty}_{\text{o}}(\textbf{c},\tau^*)$ under constraint \eqref{equ.opt.a} (called problem {\bf P3}) is
\vspace{-2mm}\begin{equation}
\label{equ.c_i_c_form}
\textstyle
{c}_i^* = \left[\frac{ i^{-\beta/2} (N^*\epsilon + 1) }{ \sum_{j=1}^{N^*}j^{-\beta/2} } -\epsilon\right]_0^1,\vspace{-2mm}
\end{equation}	
where $\epsilon \triangleq\tau^* \bar{p}_a \xi \gamma_0^{2/\alpha}$, $[x]^1_0 = \max\{\min\{x,1\},0\}$ denotes that $x$ is truncated by $0$ and $1$, $N^*$ can be obtained using bisection search from the conditions $\sum_{i=1}^{N^*} (\frac{i}{N^*})^{-\beta /2} < \frac{1}{\epsilon} + N^*$, $\sum_{i=1}^{N^*} (\frac{i}{N^*+1})^{-\beta /2}  > \frac{1}{\epsilon} + N^*$ and $N^*\leq N_f$.
\end{proposition}
\begin{IEEEproof}
The Hessian matrix of $-\underline{\mathbb{P}}^{\infty}_{\text{o}}(\textbf{c},\tau^*)$ in terms of $\textbf{c}$ can be derived as a diagonal matrix, where the $i$th element $\frac{\partial^2 -\underline{\mathbb{P}}^{\infty}_{\text{o}}(\textbf{c},\tau^*)}{\partial^2 c_i} =2q_i \frac{\tau \bar{p_a} \xi \gamma_0^{2/\alpha}}{(c_i+\tau \bar{p_a} \xi \gamma_0^{2/\alpha})^3 } >0 $. Thus, $\underline{\mathbb{P}}^{\infty}_{\text{o}}(\textbf{c},\tau^*)$ is concave in terms of $\textbf{c}$. Therefore, from the Karush-Kuhn-Tucker (KKT) conditions of problem {\bf P3}, we can obtain the optimal caching probability as
\vspace{-2mm}\begin{equation}
\label{equ.c_i_optimal}
\textstyle
{c}_i^* = \left[{\frac{x_i}{u}}- \epsilon\right]_0^1,\vspace{-2mm}
\end{equation}
where $x_i \triangleq q_i \epsilon(\epsilon+1)$ and $u$ satisfies $\sum_{i=1}^{N_f} c^*_i = 1$. Because $q_i$ is a decreasing function of $i$, ${c}_i^*$ decreases with $i$ according to \eqref{equ.c_i_optimal}. Thus, there exists a unique file index $N^*$, with which $c^*_i >0$ if $i<N^*$, and $c^*_i =0$ otherwise. As a result, finding the solution of $u$ is equivalent to finding the index $N^*$ from $\sum_{i=1}^{N^*}({\frac{x_i}{u}}- \epsilon) = 1$, which can be rewritten as $u^* = \frac{\sum_{i=1}^{N^*}x_i}{N^*\epsilon + 1}$. By substituting $u^*$ into \eqref{equ.c_i_optimal}, we can obtain \eqref{equ.c_i_c_form}. Considering $c^*_{N^*} = \frac{x_i}{u^*}- \epsilon>0$ and $c^*_{N^*+1} = \frac{x_{i+1}}{u^*}- \epsilon<0$, the conditions for $N^*$ can be obtained.
\end{IEEEproof}

According to \cite{SL.OPT}, the number of iterations of interior point method with a feasible initial point for problem \textbf{P2} is at most
$N_I = cN_{\rm ip}\lceil \sqrt{N_f+1}\log_2\frac{(N_f+1)}{\delta}\rceil$,
where $\delta$ is the tolerant accuracy, $c$ is a constant with typical value of 200 and $N_{\rm ip}$ is the number of initial values. The number of iterations of bisection method to obtain $c^*_i$ in \eqref{equ.c_i_c_form} is at most
$\lceil\log_2N_f\rceil$, which is much less than $N_I$. Therefore, the solution of problem \textbf{P3} is of low complexity.

\vspace{-5mm}
\section{Numerical and Simulation Results}\vspace{-1mm}
\label{s:5}
In this section, we validate the approximation and bound and evaluate the offloading gain of the optimized policies.

We consider a square cell with side length $500$ m. The users' locations follow a PPP with $\lambda = 0.01$, so that in average there is one user in a $10~\text{m}\times 10~\text{m}$ area. The
path-loss model is $37.6+36.8\log_{10}(r)$, where $r$ is the distance of the D2D link. $W=20$ MHz and
$\sigma^2 = -100 $ dBm, the transmit power of each DT is $P_{\rm t}=200 $ mW ($23$ dBm).
The file catalog contains $N_f=1000$ files. The parameter of the Zipf distribution $\beta=1$. This simulation setup will be used unless otherwise specified.

We consider the following caching policies for comparison.

 1) \emph{Opt. (Inter-Point)}: the solution obtained from problem \textbf{P2} by interior point method with $1000$ random initial values.

  2) \emph{Opt. (Lower-Bound)}: the low complexity solution in \eqref{equ.c_i_c_form}.

  3) \emph{Pop.}: the caching distribution is a Zipf distribution identical to the content popularity, i.e., $c_i=q_i, i=1,\cdots, N_f$.

  4) \emph{Unif.}: the caching distribution is a uniform distribution.

\vspace{-2mm}
\begin{figure}[!htb]
	\centering
	\includegraphics[width=0.45\textwidth]{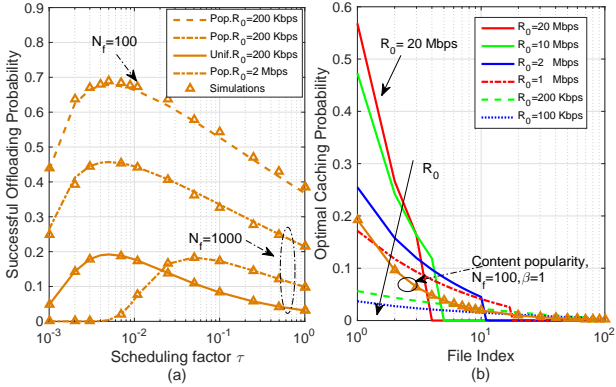}\vspace{-3mm}
	\caption{Optimal scheduling factor and caching distribution, $r_c=100$ m.}\label{fig.1}\vspace{-3mm}
\end{figure}

In Fig. \ref{fig.1}(a), we evaluate the approximation used in Proposition 1. We can see that the numerical and simulation results overlap, i.e., the approximation in \eqref{equ.t:1.1} is accurate, even when $N_f=100$ and \textbf{c} of \emph{Pop.} caching policy is skewed with $\beta=1$. In the sequel, we only show the numerical results, where noise is ignored and \eqref{equ.t:1.1} is used to compute the successful offloading probability. We can also observe that for the same data rate threshold, the optimal scheduling factors are the same for different caching distributions, which validates that $\tau^*$ is irrelevant to the caching policies. As expected, $\tau^*$ increases with the growth of $R_0$, which however is far less than 1 and larger than $1/K$ (e.g., with \emph{Pop.} for $N_f =1000$, $K=949$).


In Fig. \ref{fig.1}(b), we show the impact of data rate threshold on the caching probability, which is obtained by \emph{Opt. (Inter-Point)}. We can see that with the growth of $R_0$, the files with lower popularity have less chances to be cached and \emph{vice versa}. This is because to achieve high data rate, the distances of D2D links need to be shrunken, which makes the files with higher popularity cached with higher probability.

\vspace{-2mm}
\begin{figure}[!htb]
	\centering
	\includegraphics[width=0.45\textwidth]{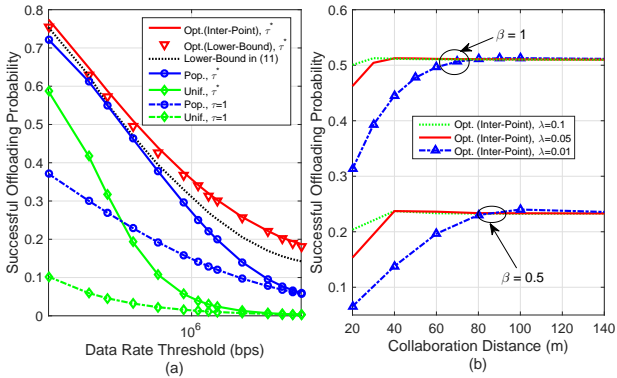}\vspace{-3mm}
	\caption{Impact of data rate threshold $R_0$ and collaboration distance $r_c$ on successful offloading probability. (a) $r_c=100$ m. (b) $R_0=200$ Kbps.}\label{fig.2}\vspace{-3mm}
\end{figure}

In Fig. \ref{fig.2}(a), we validate the lower bound in \eqref{equ_p_s_l} and show the offloading gain from joint optimization of scheduling and caching policies. As expected, the successful offloading probability decreases with the data rate threshold $R_0$. The \emph{Opt. (Inter-Point)} and \emph{Opt. (Lower-Bound)} methods can achieve almost the same performance. The successful offloading probability can be improved $200\%$ by using the joint optimization compared to \emph{Pop.} caching policy with $\tau=1$, and the gain is even larger compared to \emph{Unif.} caching policy. When $R_0$ is low, optimizing scheduling policy can introduce high gain, while when $R_0$ is high, optimizing caching policy is more important.

In Fig. \ref{fig.2}(b), we show the impact of the collaboration distance $r_c$, which plays the same role as the cluster size in \cite{Golrezaei.TWC}. We can see that with the growth of $r_c$, the successful offloading probability first increases and then almost keeps constant. This indicates that optimizing collaboration distance is unnecessary after the joint optimization of caching and scheduling policies, and $r_c$ should be large enough.


\vspace{-3mm}
\section{Conclusion}
\label{s:6}
In this letter, we jointly optimized caching and scheduling to maximize the successful offloading probability for cache-enabled D2D communications, where probabilistic
caching and random scheduling polices are considered, and both can be implemented in a distributed manner. The optimal scheduling factor with closed-form expression was first obtained, and then a local optimal caching probability was found. A low complexity caching policy was provided by maximizing a lower bound of successful offloading probability. Simulation and numerical results showed that the offloading gain can be significantly improved by jointly optimizing the content placement and delivery.


\vspace{-4mm}
\bibliographystyle{IEEEtran}
\bibliography{CBQ_J}
\end{document}